\documentclass[a4paper,11pt]{article}
\usepackage{pos}
\usepackage{enumitem}
\usepackage{hyperref}

\title{Axion Star Explosions and the Reionization History of the Universe}

\author*[a]{Miguel Escudero}

\affiliation[a]{Theoretical Physics Department, CERN,\\
  1211 Geneva 23, Switzerland\footnote{Slides can be found here \href{https://agenda.infn.it/event/34125/contributions/206411/}{https://agenda.infn.it/event/34125/contributions/206411/}}}

\emailAdd{miguel.escudero@cern.ch}

\abstract{Cosmological structure formation simulations of ultralight axion-like dark matter have shown that an axion star forms at the center of every dark matter halo in the Universe. These axion stars would then form in large numbers during the dark ages, $z \lesssim 70$. Axion stars would represent the densest axion environments in the Universe, and as such they can trigger collective processes that cannot otherwise occur for axions in vacuum. In particular, even though the lifetime of individual sub-eV axions decaying into a pair of photons is much larger than the age of the Universe, axion stars can decay into photons on very short time scales due to parametric resonance. In this talk, based on~\cite{Escudero:2023vgv,Du:2023jxh}, I will discuss the cosmological implications of such decays. We show that massive enough axion stars will decay into a large number of radio photons which will in turn lead to heating and ionization during the dark ages which is strongly constrained by Planck. As a result, we find that couplings $10^{-14}\,{\rm GeV}^{-1} \lesssim g_{a\gamma\gamma} \lesssim 10^{-10}\,{\rm GeV}^{-1}$ are excluded by Planck for $10^{-14}\,{\rm eV}\lesssim m_a\lesssim 10^{-8}\,{\rm eV}$ within our benchmark model of axion star abundance. We also highlight that future measurements of the 21 cm line can have sensitivity to couplings at least one order of magnitude smaller.}

\FullConference{1st General Meeting and 1st Training School of the COST Action COSMIC WISPers (COSMICWISPers)\\ 
5-14 September, 2023\\
Bari and Lecce, Italy\\}

\begin{document}
\maketitle

\section{Introduction: The core of the idea}

\noindent The key points behind the idea of~\cite{Escudero:2023vgv} on how axion star explosions into low-energy photons can alter the ionization history of the Universe are the following:

\begin{enumerate}[itemsep=-3pt]
    \item Cosmological simulations of ultralight axion-like dark matter matter have demonstrated that a dense solitonic core forms in the center of every dark matter halo in the Universe~\cite{Schive:2014dra,Schive:2014hza,Chan:2021bja}. 
    \item This solitonic core is what is referred to as an \textit{axion star}, and these axion stars would then represent the densest axion environments in the Universe.
    \item Occupation numbers in these axion stars are huge, and therefore they are prone to exhibit collective effects that cannot otherwise occur in vacuum. In particular, even though the lifetime of axions decaying into photons with $m_a < {\rm eV}$ is much longer than the age of the Universe, massive enough axion stars can decay via parametric resonance within days into a large number of photons with $E_\gamma =m_a/2$. This has been both discussed analytically and demonstrated numerically with explicit simulations, see e.g.~\cite{Hertzberg:2018zte,Tkachev:1987cd} and~\cite{Levkov:2020txo,Amin:2020vja}, respectively.
    \item These axion stars would form in large numbers during the dark ages $(z\lesssim 70)$ and provided that the axions that compose them interact with photons, as soon as they are massive enough they will quickly decay into a large number of photons with $E_\gamma = m_a/2$. We will be interested in $m_a < 10^{-8}\,{\rm eV}$ and these will then be radio photons. 
    \item Once axion stars decay into radio photons ($a\to \gamma \gamma$) these will be efficiently absorbed in the intergalactic medium (IGM) via inverse Bremsstrahlung, $\gamma e p \to e p$, see e.g.~\cite{Chluba:2015hma}. In turn, this will raise the temperature of the IGM, and when it becomes $T_{\rm IGM} \simeq 1\,{\rm eV}$, collisional ionization processes, $e H \to 2e p $, will lead to substantial ionization of the Universe~\cite{2010gfe..book.....M}.
    \item We know from the exquisite CMB observations by the Planck satellite that the Universe should have reionized rather late, $z_{\rm reio} \lesssim 10-12$~\cite{Planck:2016mks}. As such, an early period of cosmic reionization triggered by axion star explosions is strongly constrained by Planck data.
    \item What fraction of the dark matter would need to decay into radio photons for the Planck bound to apply? A very small one! In principle, only a fraction of $f_{\rm DM}^{\rm crit} \simeq 3\!\times\! 10^{-9}$ of the dark matter energy density that is transformed into ionizing energy will be enough to fully ionize the Universe after recombination. Why? It is very simple: energetically it only takes $E_{H}^{\rm ion}=13.6\,{\rm eV}$ to ionize a Hydrogen atom and this is only a $10^{-8}$ fraction of the proton rest mass. In addition, since the energy density of dark matter is $\simeq \! 5$ times larger than the baryon one we have $f_{\rm DM}^{\rm crit} = E_{H}^{\rm ion} n_b/\rho_{\rm DM} = E_{H}^{\rm ion}/m_p \Omega_b/\Omega_{\rm DM} \simeq 3\!\times\! 10^{-9}$. 
    In fact, this very fine sensitivity is precisely the key behind the celebrated constraint on annihilating thermal dark matter from CMB observations that tell us that s-wave annihilating WIMPs should have $m_{\rm WIMP}^{\rm s-wave} \gtrsim 10\,{\rm GeV}$, see e.g.~\cite{Slatyer:2021qgc}.
\end{enumerate}
In summary, cosmological simulations show that in axion-like dark matter cosmologies axion stars should represent a non-negligible fraction of the energy density of the Universe. In addition, these axion stars, if massive enough, can decay into a large number of radio photons. Upon decay, these radio photons would be absorbed by the intergalactic medium, and this energy injection will eventually lead to an early period of cosmic reionization. If decaying axion stars comprise only a fraction of $f_{\rm DM}^{\rm crit} \simeq 3\times 10^{-9}$ of the dark matter energy density they can lead to a period of early reionization that is strongly excluded by Planck CMB data. As such, this allows us to set bounds on the interaction strength between axions and photons ($g_{a\gamma\gamma}$) as a function of $m_a$. 

To this end, in Ref.~\cite{Escudero:2023vgv} we studied the evolution of the energy injection and subsequent reionization as triggered from these decaying axion stars by using the axion star abundance and merger rates as we first calculated in Ref.~\cite{Du:2023jxh}. In the rest of this talk, I will discuss the essence behind our analysis and also present our results. It will turn out that our phenomenology will only be relevant for axion-like dark matter particles, but to ease the nomenclature I will use the word axions throughout most of the rest of the presentation.

\section{Axion Stars in the Universe}

\noindent  Spectacular simulations of structure formation in ultra light axion-like dark matter cosmologies have shown that a dense solitonic core forms at the center of every virialized dark matter halo, see~\cite{Schive:2014dra,Schive:2014hza} for the first simulations, and~\cite{Chan:2021bja} for some of the latest ones.

These solitonic cores are called \textit{axion stars} and they are self-gravitating objects made out of non-relativistic axions, see e.g.~\cite{Visinelli:2021uve} for a recent review. Effectively, these axion stars are supported against gravitational collapse due to the uncertainty principle, and the typical de-Broglie wavelength of axions in them is comparable to the virial radius of the system, see e.g.~\cite{Hui:2016ltb}. 

The axion stars as observed in cosmological simulations form almost immediately upon halo collapse, see e.g.~\cite{Levkov:2018kau}, and they would represent the densest axion environments in the late Universe. Furthermore, the pioneering cosmological simulations of Schive et al.~\cite{Schive:2014dra,Schive:2014hza} showed that not only there is an axion star in the center of every dark matter halo, but that the mass of such an axion star ($M_S$) appeared to be strongly correlated with the mass of the dark matter halo that host it ($M_h$). In particular, the first simulations found a rather strict power law relation of the form~\cite{Schive:2014dra,Schive:2014hza}:
\begin{align}
    M_S \propto M_h^{1/3}\,\quad [{\rm Schive \, et \, al. `14}]\,.
\end{align}

Interestingly, the recent simulations of Chan et al.~\cite{Chan:2021bja} that contain a large number of simulated halos have not found a strict relation between $M_S$ and $M_h$ but rather display some diversity. Namely, for a given $M_h$ the masses of the stars seen in the center appear to be bounded to lie within a given range, although there is no observation of a strict core-halo mass relation. At this point in time it is unclear what the source of this diversity is, but what is important for our analysis is that the simulations show that the axion star mass and the halo mass are at least bounded to be within some range. In particular, all the axion stars found in the simulations of~\cite{Chan:2021bja} are found to lie within:
\begin{align}
    \qquad \qquad M_S \propto M_h^{\alpha}\,\,\,\,\, {\rm with} \,\,\,\, \alpha \in [1/3-3/5] \quad  [{\rm Chan \, et \, al. `22}]\,.
\end{align}
Namely, the latest's simulations seem to find axion stars that are heavier than those found in~\cite{Schive:2014dra,Schive:2014hza}.

The existence of one axion star at the center of every dark matter halo together with a relation between its mass and that of the halo that host it, clearly allows one to calculate the abundance of axion stars as a function of time using semi analytical methods. In particular, in~\cite{Du:2023jxh} we used the extended Press-Schechter formalism to calculate the mass function of axion stars and also their merger rate throughout cosmic history for various core-halo mass relations\footnote{Xiaolong's code is available at \href{https://github.com/Xiaolong-Du/Merger_Rate_of_Axion_Stars}{https://github.com/Xiaolong-Du/Merger\_Rate\_of\_Axion\_Stars}.}. In particular, plugging in the relevant redshift and axion mass dependencies we used the following relation:
\begin{align}
M_{S}(z) = \left[\frac{M_{h}}{M_{\rm min}(z)}\right]^{\alpha}M_{\rm min}(z)\,,
\label{eq:Core-Halo_Relation}
\end{align}
where $z$ is the redshift, $M_h$ is the virial halo mass, $\alpha$ is a power law exponent, and~\cite{Schive:2014hza}:
\begin{align}
M_{\rm min}(z)\simeq \,&1.4\times10^{-6}\left(\frac{m_a}{10^{-13}\text{ eV}}\right)^{-3/2} \, (1+z)^{3/4}M_{\odot}\,,
\label{eqn:Mmin}
\end{align}
where $m_a$ is the axion-like dark matter mass and  $M_{\rm min}$ is the smallest halo mass that could host a soliton of mass $M_S$ at a given redshift (given by the Jeans scale)~\cite{Hu:2000ke}.

\begin{figure}[t]
\centering
\includegraphics[width=0.82\textwidth]{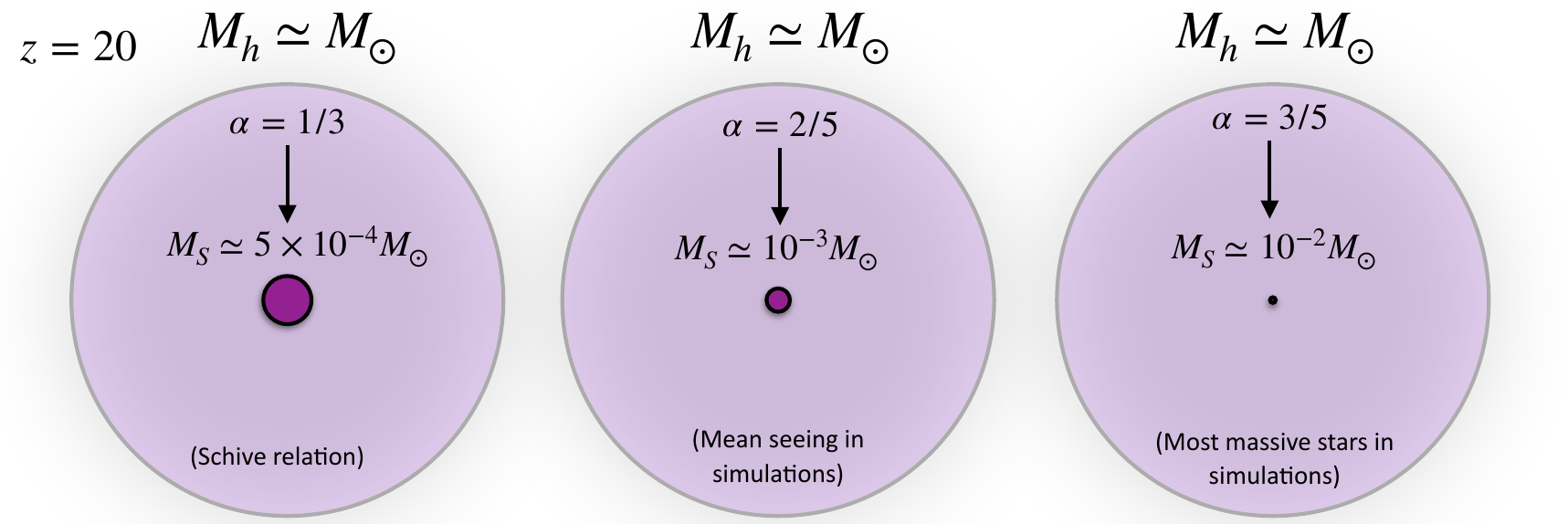}
\vspace{-0.1cm}
\caption{Pictorial representation of a $M_h = M_\odot$ dark matter halo at $z = 20$ made out of axions with $m_a = 10^{-13}\,{\rm eV}$. What is represented at the center is the axion star and its mass given the three core-halo mass relation with $\alpha = 1/3$, $2/5$, and $3/5$ in Eq.~\eqref{eq:Core-Halo_Relation}, respectively. Note that the more massive the axion star the smaller its radius, $R_S\propto 1/M_S$.} \label{fig:MsunHalo}
\end{figure}

The main uncertainty in our derived constraints arises from the present lack of detailed knowledge of core-halo mass relation. While the most recent simulations seem on average to be reproduced by $\alpha = 2/5$ there is rather large dispersion around them. For this reason, in~\cite{Escudero:2023vgv} we obtained results for $\alpha = 1/3$ (most conservative), $\alpha = 2/5$ (most realistic), $\alpha = 3/5$ (most aggressive and unrealistic). To illustrate this point, in Figure~\ref{fig:MsunHalo} I highlight a $M_h = M_\odot$ halo at $z = 20$ made out of $m_a = 10^{-13}\,{\rm eV}$ axions with its corresponding star for the three scenarios above. We can clearly notice that the choice of the power law slope parameter $\alpha$ will have a strong impact on the fraction of dark matter energy density that axion stars represent.

Figure~\ref{fig:MsunHalo} also highlights an important property of the axion stars we are interested in: they are dense compared to the cosmological mean axion density, but they are not extreme axion environments. Indeed, their compactness is very small: $C < 10^{-10}$, for the three cases depicted in the figure.

\section{Axion Stars Explosions: Parametric Resonance Decay into Photons}

Axion-like particles are generically expected to interact with the electromagnetic field via a Cherns-Simons term of the form: $\mathcal{L} = g_{a \gamma \gamma} a F^{\mu \nu}\tilde{F}_{\mu \nu}/4$. If $g_{a \gamma \gamma}\neq 0$, then axions will decay into two photons, $a \to \gamma \gamma$. In vacuum, however, for axions with $m_a < {\rm eV}$ and given the laboratory CAST limit on $g_{a\gamma\gamma}$, the axion lifetime is much longer than the age of the Universe, $\tau_a = 64\pi/(g_{a\gamma\gamma}^2m_a^3) \gg t_U$, which means that at first sight it appears very complicated to see indirect detection signatures of sub-eV decaying axion dark matter. 

However, the situation can be drastically different if the decay happens within a dense and coherent axion medium, and in particular, within an axion star. Axion stars feature two key properties in this regard: 1) they are bound solutions of axions, stabilized against collapse by a coherent field gradient, and 2) they feature huge occupation numbers, $N \simeq M_S/m_a \simeq 10^{76}\,(M_S/10^{-4}M_\odot)(10^{-13}\,{\rm eV}/m_a)$. This means that in these systems processes that do not effectively occur in vacuum can happen on short time-scales within the star.

In particular, if axions decay into photons, the phenomena of parametric resonance can take place and lead to the whole star decay into photons on much shorter time scales. The decay rate in the medium is given by $\Gamma_{\rm decay} \simeq g_{a\gamma\gamma} \sqrt{\rho_a}/2$~\cite{Levkov:2020txo}. The essence behind the idea of parametric resonance decay is simple: if only one axion of the star decays and produces two photons with $E_\gamma = m_a/2$, these photons have the right energy to in turn stimulate other axions to decay, and so on and so forth leading to a exponentially fast decay of the whole system. Namely, the system is stimulated with precisely the right frequency at which it resonates. Analytically, see e.g.~\cite{Hertzberg:2018zte,Tkachev:1987cd}, the condition for parametric resonance decay can be approximately written as the requirement that once produced, these photons should be able to at least stimulate another axion within the system:
\begin{align}
    \Gamma_{\rm decay} \times R_S  \simeq g_{a\gamma\gamma} \sqrt{\rho_a} \times R_S  > 1 \, \quad \text{[parametric resonance condition]}\,,
\end{align}
where here $R_S$ is the radius of the axion star. Indeed, axion star explosions have been seen in explicit numerical simulations fulfilling this requirement, and in particular in two cases: 1) for axion stars that are above this critical threshold, as well as 2) those generated by mergers of sub-critical axion stars of similar mass~\cite{Levkov:2020txo,Amin:2020vja}. In particular, Ref.~\cite{Levkov:2020txo} showed that axion stars above the following mass threshold will be unstable and decay:
\begin{equation}\label{eq:Mdecay}
M_{S}^{\rm decay}\simeq 8.4\times 10^{-5}\,M_\odot\left(\frac{10^{-11}\,\text{GeV}^{-1}}{g_{a\gamma\gamma}}\right) \left( \frac{10^{-13}\,\text{eV}}{m_a}\right) \,.
\end{equation}
The decay lifetime simply corresponds to the light crossing time within the star and reads:
\begin{align}\label{eq:taudecay}
\tau_{S}^{{\rm decay}} \simeq r_c \simeq {\rm day} \left(\frac{8.4 \times 10^{-5} M_\odot}{M_S} \right) \left(\frac{10^{-13}\,{\rm eV}}{m_a}\right)^2\,.
\end{align}

As seen in Figure~\ref{fig:MsunHalo}, for $m_a= 10^{-13}\,{\rm eV}$ and $g_{a\gamma \gamma} > 10^{-12}{\rm GeV}^{-1}$ all halos with $M_h > M_\odot$ will host a star that will be unstable and decay on a very short time scale. Since we expect many halos in the Universe with $M_h > M_\odot$ it is clear that the critical axion star abundance could be larger than $f_{\rm DM}^{\rm decay} \simeq 10^{-9}$ and CMB constraints will indeed apply.

\section{Abundance of Critical Axion Stars, Merger Rates, and Plasma Effects}

\begin{figure}[t]
\centering
\begin{tabular}{cc}
\hspace{-0.3cm}\includegraphics[width=0.48\textwidth]{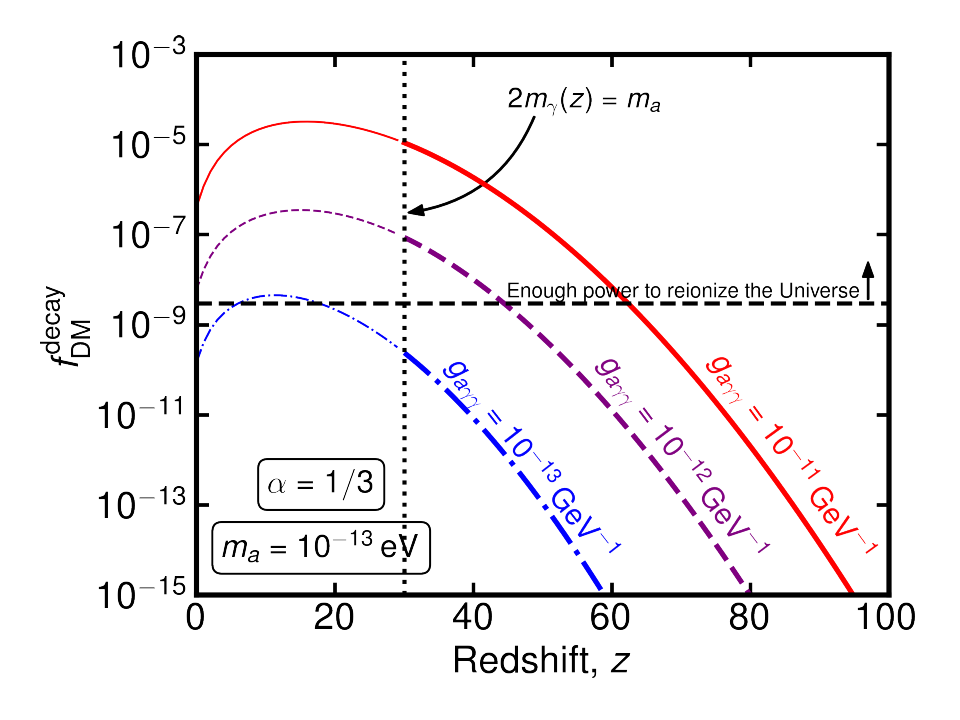} & \includegraphics[width=0.48\textwidth]{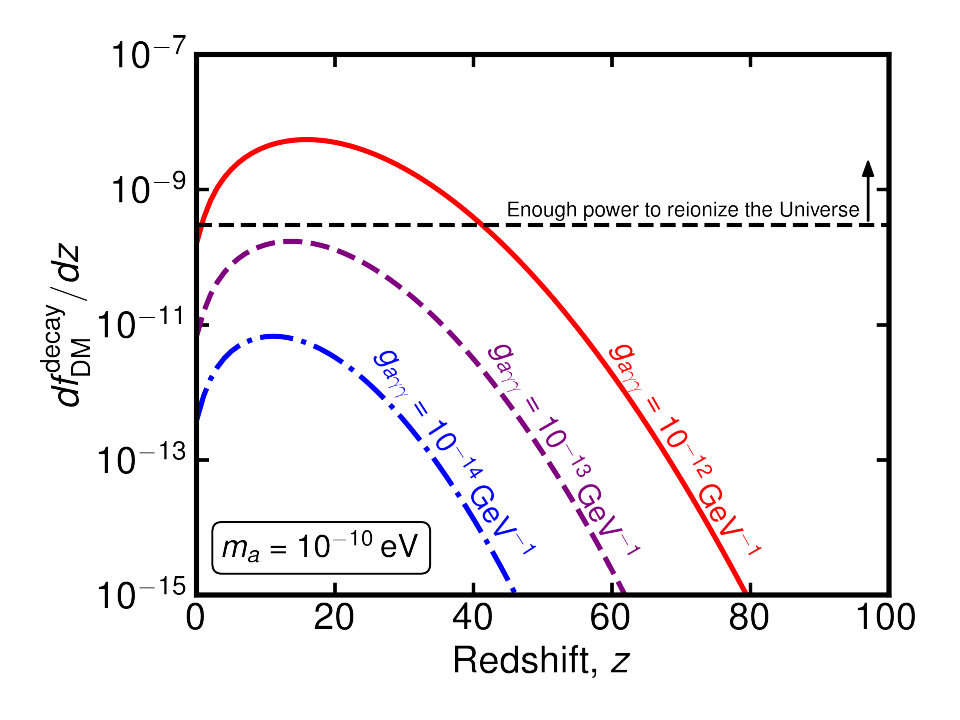}
\end{tabular}
\vspace{-0.5cm}
\caption{\textit{Left:} Fraction of dark matter in the form of critical axion stars for $m_a = 10^{-13}\,{\rm eV}$ and $\alpha = 1/3$. These critical axion stars will all explode at $z \simeq 30$ because then $m_\gamma(z) < m_a/2$. \textit{Right:} Continuous energy injection rate $(df_{\rm DM}/dz)$ of dark matter injected into low energy photons as a result of mergers of sub-critical axion stars for $m_a = 10^{-10}\,{\rm eV}$ and a few benchmark examples for $g_{a\gamma\gamma}$. Both figures were obtained from the results of Ref.~\cite{Du:2023jxh}, see also footnote 1.} \label{fig:energydensity}
\end{figure}

As we have seen, axion stars with a mass above the threshold in Eq.~\eqref{eq:Mdecay} will be unstable and decay on a very short timescale into photons with $E_\gamma = m_a/2$. Effectively, the energy emitted in the decay will be $E \simeq M_S - M_S^{\rm decay}$~\cite{Levkov:2020txo}. Namely, the decay process would always leave a remnant star with a mass close to the critical decay threshold. It is clear then that as soon as an axion star mass crosses the threshold for decay it will do so. This means that we expect most of the decays to happen as soon as the axion stars form in large numbers and grow massive enough in the early Universe.

In the scenario that we consider, and as done in the numerical simulations~\cite{Schive:2014dra,Schive:2014hza,Chan:2021bja}, the initial power spectrum of perturbations is almost scale invariant and adiabatic, and structure formation happens hierarchically. As such, axion stars will form upon collapse of the halos that host them. Then, via Eq.~\eqref{eq:Core-Halo_Relation} we can track the mass evolution of these stars, and we will be able to know what fraction of dark matter mass there is in critical axion stars at a given redshift by following the mass function of dark matter halos that host them. In~\cite{Du:2023jxh}, we used the extended Press-Schechter formalism to calculate precisely this. In the left panel of Figure~\ref{fig:energydensity}, we show the fraction of energy density in critical axion stars for $m_a = 10^{-13} \,{\rm eV}$ for the axion star abundance model of $\alpha = 1/3$. We can clearly see that the fraction only becomes sizeable at $z \lesssim 70$, during the dark ages. 

Of course, as soon as the threshold for decay in Eq.~\eqref{eq:Core-Halo_Relation} is met, all critical axion stars in the Universe should decay into photons. However, this may not be entirely true. In the early Universe, there are always some free charged particles, and this means that the photon is not massless, but rather displies a finite plasma mass. In particular, $m_\gamma(z) > 10^{-14}\,{\rm eV}$, see e.g.~\cite{Raffelt:1996wa}. Photons with smaller masses than this cannot propagate and hence axions with $m_a < 2m_\gamma(z)$ cannot actually decay into photons. The photon plasma mass depends upon the density of charged particles in the plasma and cosmologically it varies between $m_\gamma \simeq 2\times 10^{-13}\,{\rm eV}$ at $z=100$ to  $m_\gamma \simeq 10^{-14}\,{\rm eV}$ right before standard cosmic reionization occurs. This means that there cannot be any axion star explosions into photons if $m_a \lesssim 2\times 10^{-14}\,{\rm eV}$, and that for $2\times 10^{-14}\,{\rm eV} \lesssim m_a \lesssim 10^{-12}\,{\rm eV}$ axion stars could be much heavier than the threshold in Eq.~\eqref{eq:Mdecay}, but still not decay due to plasma blocking effects. 

For $m_a \gtrsim 10^{-12}\,{\rm eV}$, axion stars will decay as soon as they pass the threshold in Eq.~\eqref{eq:Mdecay} and leave a remnant critical axion star in the center of the halo. Would this mean that then we will not expect more decays? The answer is no. We expect further decays to happen via mergers of these axion stars as hierarchical structure formation proceeds and halos that host critical axion stars undergo major mergers. Once the two critical stars in each of the individual halos merge, an energy of $E\sim M_S^{\rm crit}$ will be release in photons with $E_\gamma = m_a/2$~\cite{Levkov:2020txo,Amin:2020vja}. In Ref.~\cite{Du:2023jxh} we calculated the rate at which this happens using the extended Press-Schechter model to semi analytically evaluate the energy density released in photons with $E_\gamma = m_a/2$, and we explicitly checked that it agreed against results we obtained from stochastic merger trees. The result of this merger rate is shown in the right panel of Figure~\ref{fig:energydensity}. We again clearly see how this rate can become significant at $z \lesssim 70$.

\section{Cosmological Impact of Radio Photon Energy Injections from Axion Star Explosions}

Axion star explosions will lead to large energy releases of very low frequency photons, $E_\gamma =m_a/2$. The most efficient absorption process of these photons is via inverse Bremsstrahlung on pairs of free electrons and protons, $\gamma e p \to e p$~\cite{Chluba:2015hma}. The rate for absorption depends strongly on $E_\gamma$ as well as on the density of $e/p$, but for $m_a \lesssim 10^{-8}\,{\rm eV}$ the absorption length scale is always shorter than the size of the observable Universe at the time when these photons are produced. Since the phenomenology depends upon the photon absorption length-scale we separated it in two regimes: 
\begin{enumerate}
    \item $m_a\lesssim 10^{-13}\,{\rm eV}$ -- In this regime, the photons produced from axion star explosions are absorbed on very small lengthscales and we treat the explosions as shockwaves in the IGM using similar methods to those used for supernova explosions. This will lead to patchy reionization.
    \item $m_a\gtrsim 10^{-13}\,{\rm eV}$ -- In this regime, the photons are absorbed over lengthscales typically larger than the inter-separation between axion stars that explode, and as such we can treat the process as homogeneous throughout the whole Universe. 
\end{enumerate}

To track the baryon temperature as well as the free electron fraction of the Universe we consider all the relevant cooling and heating processes: 1) heating caused by axion star explosions into photons as absorbed by inverse Bremsstrahlung, 2) adiabatic cooling (expansion of the Universe), 3) Compton cooling ($e \gamma_{\rm CMB} \to e \gamma$), 4) collisional excitation cooling ($eH \to e H^* \to e H \gamma$), and 5) collisional ionization cooling ($eH\to 2 ep$). We then track the free electron fraction in the Universe by considering 1) recombination ($ep\to H \gamma$), 2) photoionization ($\gamma H \to e p$), and 3) collisional ionization processes ($eH\to 2 ep$).

Taking into account all these processes we track the temperature of baryons in the Universe ($T_{\rm IGM}$) and the free electron fraction ($x_e$) using an effective 3-level atom. Then, we calculate the integrated optical width of CMB photons from reionization and contrast it against the results inferred from Planck CMB observations~\cite{Planck:2016mks}. In practice, since we know the Universe should be fully ionized by $z \sim 6$, we consider a possible non-standard reionization history only for $z > 6$. We note that this is clearly conservative as considering standard reionization (by normal stars) on top of that generated by axion stars will lead to stronger constraints.

\section{Results}

\begin{figure}[t]
\centering
\begin{tabular}{cc}
\hspace{-0.3cm}\includegraphics[width=0.49\textwidth]{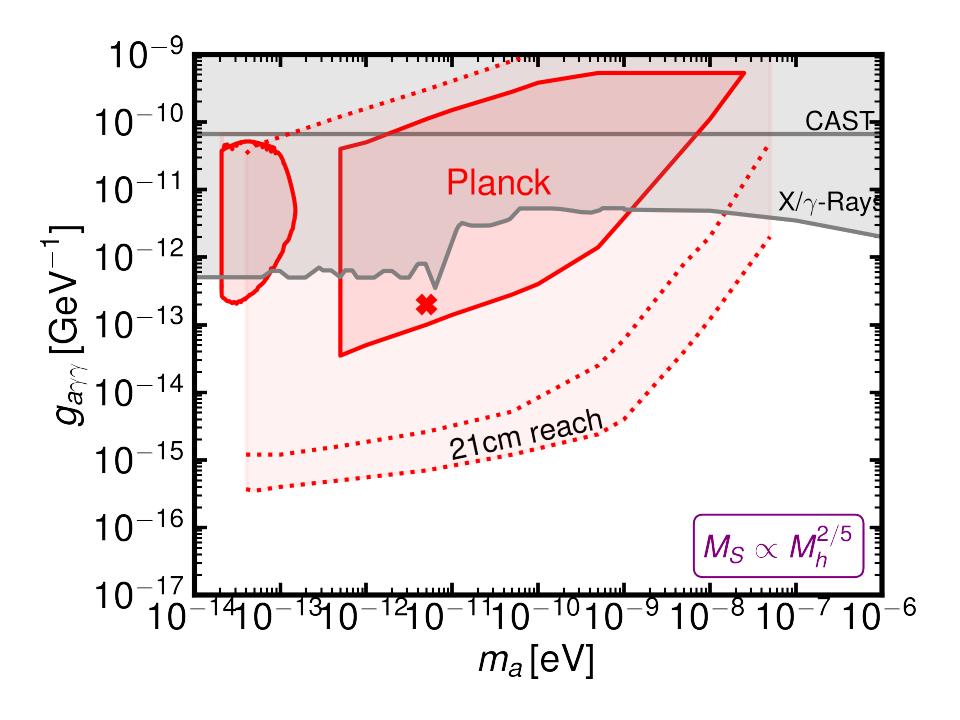} & \includegraphics[width=0.45\textwidth]{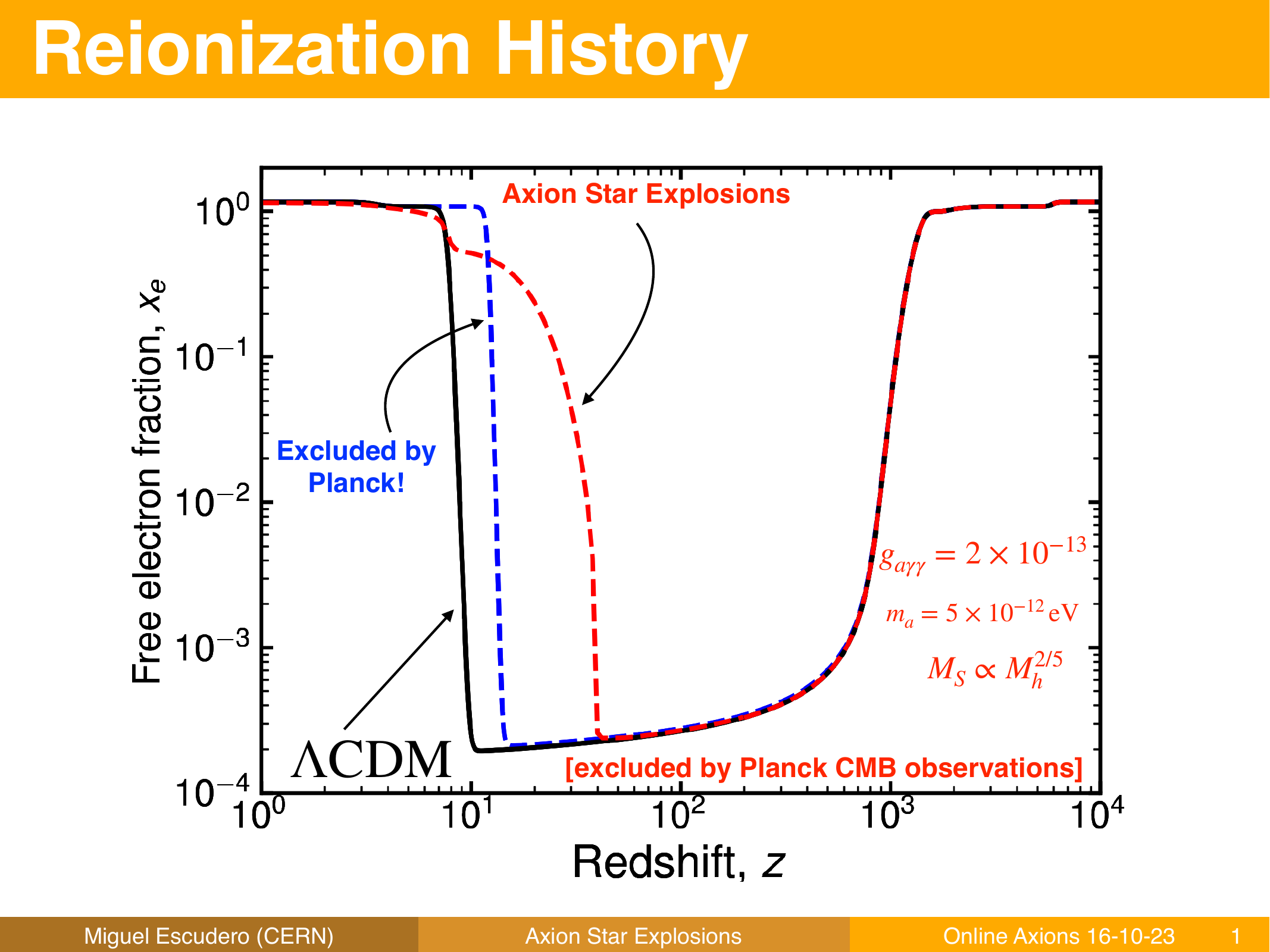}
\end{tabular}
\vspace{-0.4cm}
\caption{\textit{Left:} Parameter space of $g_{a\gamma\gamma}$ coupling as excluded by Planck CMB observations for an axion star abundance model with $\alpha = 2/5$~\cite{Escudero:2023vgv,Du:2023jxh}. \textit{Right:} Evolution of the free electron fraction for the \textbf{\color{red}{x}} benchmark example in the left panel, and in $\Lambda$CDM for $\tau_{\rm reio} = 0.05$ (black, allowed) and for $\tau_{\rm reio} = 0.1$ (blue, excluded). }\label{fig:parameterspace}
\end{figure}

In the left panel of Figure~\ref{fig:parameterspace}, I show our resulting Planck constraints for our benchmark axion star abundance model with $\alpha = 2/5$. Across this region of parameter space axion star explosions lead to an early period of cosmic reionization that is not compatible with Planck CMB observations of the Thomson optical width ($\tau_{\rm reio}$). In the right panel, I display the evolution of the free electron fraction as a function of redshift for the example highlighted in the left panel. 

Each of the regions are explained with great detail in Ref.~\cite{Escudero:2023vgv}, but effectively they can be understood as follows: 1) the bounds do not extend up to arbitrarily high $g_{a\gamma\gamma}$. This happens because the abundance of stars that would be critical for such values will be suppressed because these stars would be hosted in halos with a mass smaller than the Jeans scale and would not have formed in the first place, see Eq.~\eqref{eqn:Mmin}. 2) The region with $m_a \simeq 3\times 10^{-13}\,{\rm eV}$ between the two disjoint exclusion regions is not excluded by Planck because here the photon plasma mass is very similar to $m_a/2$, and as soon as some reionization happens, the plasma mass grows and further axion star decays are blocked and cannot occur any more. 3) The excluded region around $m_a \simeq 5\times 10^{-14}\,{\rm eV}$ corresponds to axion star explosions of which its energy is absorbed on small length scales. This leads to patchy reionization that is not so effectively constrained by Planck. 4) The main contour extending from $m_a\simeq 10^{-12}\,{\rm eV}$ up to $m_a \simeq 10^{-8}\,{\rm eV}$ corresponds to bounds resulting from continuous axion star explosions as generated by mergers of critical axion stars. Up to $m_a \simeq 10^{-9}\,{\rm eV}$ it contains regions of parameter space where $df_{\rm DM}/dt/\tau_{\rm Compton} \simeq 3\times 10^{-9}$ (see Figure 7b of~\cite{Escudero:2023vgv}). This is indeed what we expected energetically: it corresponds to fractions of the dark matter energy density of $\simeq 10^{-9}$ that are injected over timescales shorter than the fastest cooling timescale, in this case Compton cooling. Finally, the region for $m_a \gtrsim 10^{-9}$ changes the slope because for these masses the photons from axion star explosions start to have absorption length-scales comparable to the size of the observable Universe. Eventually, the photons are not effectively absorbed and the bound stops at $m_a\simeq 2\times 10^{-8}\,{\rm eV}$.

\section{Summary and Conclusions}

Cosmological simulations of axion-like dark matter cosmologies have shown that there should be an axion star at the center of every dark matter halo in the Universe. These axion stars will represent the densest axion environments in the late Universe and due to their coherence and large axion occupation numbers they can trigger collective effects that are not possible in vacuum. 

In particular, numerical simulations have shown that if axions interact with $F\tilde{F}$, then axion stars above a certain critical threshold will be unstable and will decay fast into a huge number of radio photons with $E_\gamma = m_a/2$. 

In Ref.~\cite{Du:2023jxh} we have calculated the abundance of these critical axion stars as well as their merger rate, and in Ref.~\cite{Escudero:2023vgv} we have tracked the cosmological evolution of the energy release of these stars. 

We have shown that axion stars explosions into photons can lead to an early period of cosmic reionization and that this is constrained by Planck so long as a fraction of $f_{\rm DM}^{\rm crit} \sim 10^{-9}$ of the dark matter energy density is released by axion star explosions into photons. This has allowed us to set constraints on the $g_{a\gamma\gamma}$ coupling as a function of $m_a$ for certain axion-like dark matter models. 

\section{Outlook: Numerical Simulations, 21 cm Cosmology, and Axion Model Dependency}

\noindent To conclude, I would like to finish this presentation by highlighting three further important points:

\begin{enumerate}
    \item \textit{Need of further Numerical Simulations of Axion Stars -- } The bounds that we derive are strongly dependent upon the relation between the mass of the halo and the axion star that sits at its center. The latest simulations show that given a halo mass, the mass of the star should be bounded from above and below but that it displays diversity~\cite{Chan:2021bja}. The origin of this diversity is not clear and further studies are needed to understand the issue. As shown in Figure~\ref{fig:parameterspace_corehalomass}, this has a strong impact in the context of our study, as bounds on $g_{a\gamma\gamma}$ can move orders of magnitude up or down depending upon the $M_S(M_h)$ relation. Although we have used $\alpha = 2/5$ as a benchmark because it reproduces well the diversity seen in simulations, understanding the underlying star mass distribution would be needed to establish fully robust constraints.

\begin{figure}[t]
\centering
\includegraphics[width=0.55\textwidth]{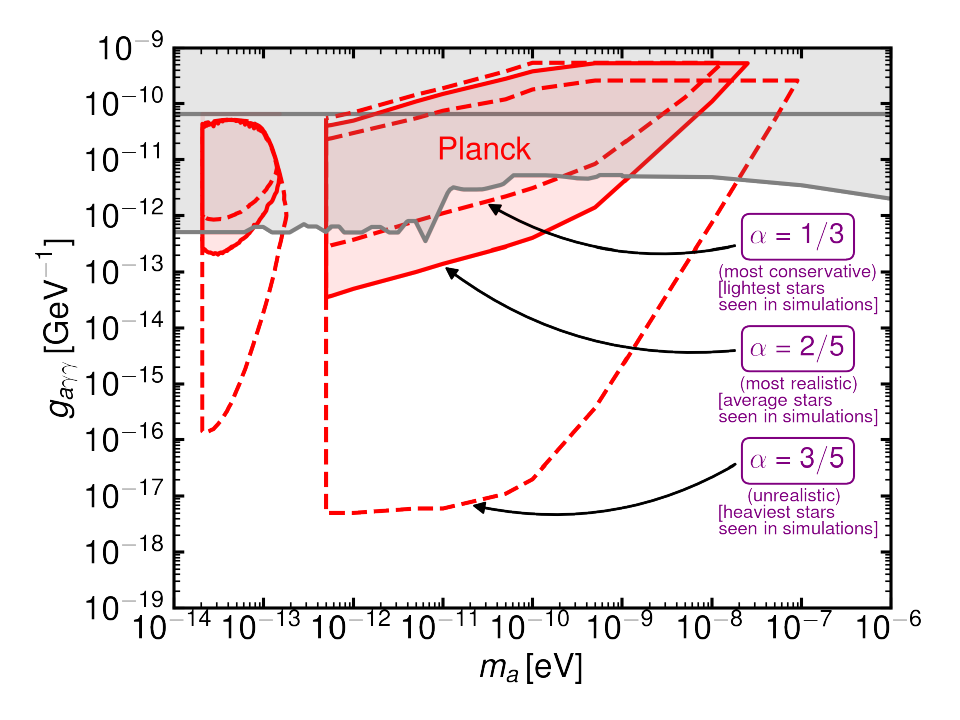} 
\vspace{-0.4cm}
\caption{Dependence of the Planck constraints upon the assumed core-halo mass relation~\cite{Escudero:2023vgv}. Our benchmark case corresponds to the case with $\alpha = 2/5$ as it describes well the diversity seen in the latest cosmological simulations of~\cite{Chan:2021bja}. The case $\alpha = 1/3$ corresponds to the lowest mass axion stars seen in cosmological simulations and is therefore conservative, while the case $\alpha = 3/5$ corresponds to the most massive axion stars seen in simulations and is therefore quite unrealistic.}\label{fig:parameterspace_corehalomass}
\end{figure}

    Furthermore, the cosmological simulations of~\cite{Schive:2014dra,Schive:2014hza,Chan:2021bja} are performed with $m_a \sim 10^{-22}\,{\rm eV}$. The Schrodinger-Poisson equations solved there features a scaling symmetry that allows in principle extrapolation to any axion mass, see e.g.~\cite{Chan:2021bja}. However, it would nevertheless be key to explicitly check the evolution of axion stars and their mergers in the mass regime of interest for our study $10^{-14}\,{\rm eV}\lesssim  m_a \lesssim 10^{-6}\,{\rm eV}$. 

    \item \textit{21cm Cosmology -- } The main effect of axion star explosions in the early Universe is to heat up the intergalactic medium. If the heating is large enough as to raise the baryon temperature to $T_{\rm IGM} \sim 1\,{\rm eV}$, then reionization can occur, and this is constrained by Planck CMB observations via measurements of the Thomson optical width. However, if one were to be able to observe directly the thermal state of the IGM, the sensitivity to axion star explosions could be much greater. This is indeed what future measurements of the 21cm line are expected to achieve, and as highlighted in Figure~\ref{fig:parameterspace} they can improve sensitivity by orders of magnitude in $g_{a\gamma\gamma}$. In order to obtain this sensitivity we performed a very simple global $T_{21}$ estimate, but it would be very interesting to perform a full simulation and obtain the power spectrum as well. It would also be interesting to look at other signatures, such as heating of the Lyman-$\alpha$ forest, or the effects of axion star decays in other lower frequency cosmic microwave backgrounds. 

    \item  \textit{Axion Model Dependence -- } We considered that axion stars decay into photons as triggered by a non-zero $g_{a\gamma\gamma}$ coupling. However, since axion stars feature such huge occupation numbers other tiny axion couplings could be relevant in the axion star evolution. Indeed, if axions have attractive self-interactions, axion stars have also been shown to become unstable and decay into relativistic axions above a mass threshold given by $M_S^{\rm Bosenova} \simeq 12.4 M_{\rm Pl}/\sqrt{|\lambda|}$~\cite{Levkov:2016rkk,Eby:2015hyx}. This should be compared with Eq.~\eqref{eq:Mdecay}, and the comparison highlights that for $\lambda = m_a^2/f_a^2$ unless $g_{a\gamma\gamma} \gtrsim 600\times \alpha_{\rm EM}/(2\pi f_a)$, the axion stars will actually decay into relativistic axions and not into photons. This clearly means that the bounds of Figure~\ref{fig:parameterspace} only apply to models with either enhanced $g_{a\gamma\gamma}$ couplings or reduced quartic interactions. This is not something particular, as bounds on $g_{a\gamma\gamma}$ significantly above the QCD axion line indeed only apply to such types of scenarios. In this context, there are many interesting models in the literature that feature such enhanced $g_{a\gamma\gamma}$ couplings and our constraints bound them, see e.g.~\cite{Farina:2016tgd,Agrawal:2017cmd,DiLuzio:2017pfr,Daido:2018dmu,Choi:2020rgn,Plakkot:2021xyx,Sokolov:2022fvs}.
\end{enumerate}

\noindent \textbf{Acknowledgements:} It is a pleasure to thank my collaborators on axion star explosions: Charis Pooni, Malcolm Fairbairn, Diego Blas, Xiaolong Du, and Doddy Marsh. Great thanks for the hard work, key ideas, and good spirit that made our papers (and hence this contribution) possible. Thanks also go to the organizers of the 1st COSMICWISPers workshop for their very kind invitation and for setting up such an enjoyable and interesting meeting. I would also like to thank Jordi Miralda Escud\'{e} for interesting discussions on axion star formation.

\setlength\parskip{-0.5pt}
\setlength\baselineskip{0pt}
\setlength\itemsep{-10pt}

\bibliography{biblio}

\end{document}